\documentclass[reqno]{amsart} \usepackage{amscd}
\usepackage{epsf}
\newtheorem{theorem}{Theorem}[section]
\newtheorem{proposition}[theorem]{Proposition}
\newtheorem{lemma}[theorem]{Lemma}

\theoremstyle{remark} 
\numberwithin{equation}{section}
\newcommand{\field}[1]{\ensuremath{\mathbb{#1}}}
\newcommand{\CC}{\field{C}}

\begin{document}
\title[ HyperK\"{a}hler Prequantization ]{HyperK\"{a}hler prequantization of 
 the Hitchin system and Chern-Simons theory with complex gauge group}
\author{Rukmini Dey}

\maketitle

\begin{abstract}
Hitchin has shown that the moduli space ${\mathcal M}$ of the dimensionally 
reduced self-dual Yang-Mills equations has a hyperK\"{a}hler structure. 
In this paper we first explicitly show the hyperK\"{a}hler structure, the 
details of which 
is missing in Hitchin's paper. We show here
that ${\mathcal M}$ admits three  pre-quantum line bundles,
corresponding to the  three  symplectic forms. We use Quillen's 
determinant
 line bundle 
construction and show that the Quillen curvatures of these prequantum line 
bundles are proportional to each of the symplectic forms mentioned above.
The prequantum line bundles are holomorphic with respect to their 
respective complex structures. We show how these prequantum line bundles can 
be derived from cocycle line bundles of Chern-Simons gauge theory with complex 
gauge group in the case when the moduli space is smooth. 
\end{abstract}         
\maketitle


\medskip

arxiv address: math-ph/0605027

\medskip

\section{Introduction}

Given a symplectic manifold $({\mathcal M}, \Omega)$ geometric 
prequantization  is a construction of a prequantum line bundle ${\mathcal L}$ on ${\mathcal M},$ whose curvature is proportional to the symplectic form.
If ${\mathcal M}$ admits such a prequantum line bundle ${\mathcal L}$, then 
one can associate a Hilbert space, namely, the square integrable sections of 
${\mathcal L}$ and a correspondence between functions on ${\mathcal M}$ to 
operators acting on the Hilbert space such that the Poisson bracket of 
two functions corresponds to the commutator of the corresponding operators. 
 The latter is ensured by the fact that 
the curvature of the prequantum line bundle is precisely 
the symplectic form $\Omega$ ~\cite{Wo}. 
Let $f \in C^{\infty}({\mathcal M})$. Let $X_f$  be the vector field 
defined by $\Omega(X_f, \cdot) = - df$. Let $\theta$ be the symplectic 
potential correponding to $\Omega$. Then we can define the operator 
corresponding to the function $f$ to be 
$\hat{f} = -i \hbar [X_f - \frac{i}{\hbar}\theta(X_f)] + f$. 
Then if $f_1, f_2 \in C^{\infty}({\mathcal M})$ and 
$f_3 = \{ f_1, f_2 \}$, Poisson bracket of the two induced by the symplectic 
form, then $[\hat{f}_1, \hat{f}_2] = -i \hbar \hat{f}_3$. 
When ${\mathcal M}$ has hyperK\"{a}hler structure, there are three symplectic 
structures and hence three Possion brackets and one can often construct 
prequantum line bundles for each one of them. This is called hyperK\"{a}hler
prequantization.

 A relevant example  in our context would be geometric
quantization of the  moduli space of flat connections on a principal 
$G$-bundle $P$ on a compact Riemann surface $\Sigma$, ~\cite{Wi}, ~\cite{ADW}. 
We describe this in some detail. 
Let ${\mathcal A}$ be 
the space of Lie-algebra valued connections on the principal bundle $P$. Let 
${\mathcal N}$ be the moduli space of flat connections (i.e. the space of 
flat connections modulo the gauge group).  
The prequantum line bundle is the  the determinant line bundle of the Cauchy-Riemann 
operator, namely, ${\mathcal L}
= \wedge^{\rm{top}} ({\rm Ker} \bar{\partial}_A)^{*} \otimes \wedge^{\rm{top}}({\rm Coker}
\bar{\partial}_A)$. It carries 
the Quillen metric such that the canonical unitary connection  has a 
curvature form which coincides with the natural K\"{a}hler form on the 
moduli space of flat connections on vector bundles over the Riemann surface
 of a given rank ~\cite{Q}. To elaborate, on the affine space of all connections, there is a natural symplectic form, proportional to 
${\rm Tr} \int_{\Sigma} \alpha \wedge \beta$, 
where $\alpha, \beta \in T_{A} {\mathcal A} = \Omega^{1}(M, {\rm ad} P).$
It can be shown, using a moment map construction, that this symplectic form
descends to the moduli space of flat connections.
One can show that the determinant line bundle equipped with the Quillen metric 
has curvature proportional to this symplectic form ~\cite{Q}. 

Inspired by this construction, we constructed a 
prequantum line bundle on the 
moduli space of  solutions to the vortex equations ~\cite{D1}. In this paper 
 geometrically quantize the hyperK\"{a}hler structure in the Hitchin system.
We elaborate on this.

The self-duality equations on a Riemann surface arise from dimensional 
reduction of self-dual Yang-Mills equations from $4$ to $2$ 
dimensions ~\cite{H}.  
They have been  studied 
extensively in ~\cite{H}. They are as follows. Let $M$ be a compact Riemann 
surface of genus $g>1$ and let $P$ be a principal 
$U(n)$-bundle over $M$. Let $A$ be a unitary connection on $P$, 
i.e. $A = A^{(1,0)} + A^{(0,1)}$ such that 
$A^{(1,0)} = -A^{(0,1)*}$, where $*$ denotes conjugate transpose 
~\cite{GH},~\cite{K}. Thus we can identify the space of all unitary 
connections with its $(0, 1)$-part, i.e. $A^{(0,1)}.$ 
Let $\Phi^{1,0}$ be a complex Higgs field, such that
$\Phi^{1,0} \in {\mathcal H} = \Omega^{1,0}(M; {\rm ad} P \otimes \CC)$. 

{\bf Note:} In ~\cite{H} this $\Phi^{1,0}$ is written as $\Phi$. But we will 
be using the present notation since we will define $\Phi^{0,1}$ as well and our
$\Phi = \Phi^{1,0} + \Phi ^{0,1}.$ 
 
The pair $(A, \Phi^{1,0})$ will be said to satisfy the self-duality equations if
$$(1)\rm{\;\;\;\;\;}F(A) = -[\Phi^{(1,0)}, \Phi^{(1,0)*}],$$
$$(2)\rm{\;\;\;\;\;}d^{\prime\prime}_A \Phi^{(1,0)} = 0.$$ 
Here $F(A)$ is the curvature of the connection $A.$
The operator $d_A^{\prime \prime}$ is the $(0,1)$ part of the extension of the 
covariant derivative operator to act on $\Omega^{1,0}(M, {\rm ad} P \otimes \mathbb{C})$.  Also $\Phi^{(1,0)*} = \phi^* d \bar{z}$ where $\phi^*$ is taking 
conjugate transpose of the matrix of $\phi$.
There is a gauge group acting on the space of $(A, \Phi)$ which leave the 
equations invariant. If $g$ is an $U(n)$ gauge transformation then 
$(A_1, \Phi_1)$ and $(A_2, \Phi_2)$ are gauge equivalent if 
$d_{A_2} g = g d_{A_1}$ and $\Phi_2 g = g \Phi_1$ ~\cite{H}, page 69.
  Taking the quotient by the gauge group of the solution 
space to $(1)$ and $(2)$ gives  the moduli space of solutions to these 
equations and is denoted by ${\mathcal M}$.
Hitchin shows that there is a natural metric on the moduli space 
${\mathcal M}$ and further  proves that the metric is 
hyperK\"{a}hler ~\cite{H}.

This paper is a sequel of the paper ~\cite{D} where we constructed 
the prequantum line bundle on ${\mathcal M}$ whose curvature is  the first 
symplectic 
form of ~\cite{H}. In ~\cite{D} we had explicitly given this symplectic form , the metric and the moment map construction (details of which are missing in ~\cite{H}). But the prequantum 
line bundle we constructed in ~\cite{D} is a bit unnatural since we used 
$\overline{\partial} + \overline{A_0^{(1,0)}} 
+ \overline{\Phi^{(1,0)}}$ which gauge transforms like  $\bar{g} (\overline{\partial} + \overline{A_0^{(1,0)}} 
+ \overline{\Phi^{(1,0)}})\bar{g}^{-1}.$ In this paper, 
we first rectify this and construct the prequantum bundle on 
${\mathcal M}$ corresponding to 
 the first symplectic form using $\bar{\partial} + A_0^{(0,1)} + \Phi^{(0,1)}$
which gauge transforms like $g(\bar{\partial} + A_0^{(0,1)} + \Phi^{(0,1)})g^{-1}$ which is more natural to use.
Next, in this paper, we construct the prequantum line
  bundles on ${\mathcal M}$ corresponding to the other two symplectic forms 
which give rise to the hyperK\"{a}hler structure.
 We show the metric, the three symplectic forms, the three complex 
structures and the three prequantum lines bundles explicitly. 
In the next section we discuss the holomorphicity of the prequantum line 
bundles w.r.t. the three complex structures.  In the last section we show  how the prequantum line bundles are related to cocycle line bundles in Chern-Simons 
gauge theory with complex gauge group at least in the case when the moduli space is smooth.  In ~\cite{Wi1} Witten had quantized  one of the symplectic forms from Chern-Simons gauge theory with complex gauge group. We find that by introducing a parameter $\lambda$ we can obtain all three 
symplectic forms from this theory -- though applications
of this as in ~\cite{G} is still a topic of  research.

Papers which may be of interest in this context are ~\cite{BS}, ~\cite{BDr},~\cite{BDr1}, ~\cite{H1}, ~\cite{S} ~\cite{Wi1}. These papers use  algebraic geometry and 
algebraic topology and may provide alternative methods to quantizing the 
hyperK\"{a}hler system, though ours is the only paper we have seen 
in which all three quantizations appear explicitly.  Our method  is very 
elementary and we 
explictly construct 
the prequantum line bundles. The only machinery we use is Quillen's  
construction of the determinant line bundle, ~\cite{Q}. It would be 
interesting to see if there is any relation between the present quantizations 
and the ones appearing in the previous papers.

 After writing the paper the author found  Kapustin and Witten's paper
~\cite{KW} where they have applied Beilinson and Drinfeld's quantization of one of the symplectic forms 
of the Hitchin system  to study the geometric 
Langlands programme. We should mention that though the metric on the 
moduli space is exactly the 
same their complex structure ${\mathcal J}$  is our $-{\mathcal K}$ and 
their ${\mathcal K}$ is our ${\mathcal J}$. In the end of section $2$, we will 
explain the dictionary between the physicists' notation for the symplectic 
forms in ~\cite{KW} and our notation.   
It would be interesting to see if there is any relevance of all the three 
quantizations obtained here to geometric Langlands programme. 

Geometric prequantization of the moduli spaces of vector bundles (with fixed determinant)  over a Riemann surface can be found in ~\cite{TZ1}. 
Some  interesting applications of the determinant line bundles  to geometry 
and physics can be found in ~\cite{TZ2}, ~\cite{TZ3}.

\section{Symplectic and hyperK\"{a}hler  structures}

Let the configuration space be defined as ${\mathcal C} = \{ (A^{0,1}, \Phi^{1,0})| 
A^{0,1} \in {\mathcal A}, \Phi^{1,0} \in {\mathcal H} \}$ where
${\mathcal A}$ is the space of unitary connections on $P$, identified with its $A^{(0,1)}$ part  
and 
${\mathcal H} = \Omega^{(1,0)}(M, {\rm ad} P \otimes \CC)$ is the space 
of Higgs field. We can extend the Higgs field $\Phi^{(1,0)}$ by its $(0,1)$ 
part by defining $\Phi^{(0,1)} = -\Phi^{(1,0)*}$, where $*$ is conjugate 
transpose. But only the $(1,0)$ parts 
belong to  ${\mathcal H}$ and appear in eqn $(1)$ and $(2)$. 
   Unitary connections satisfy $A = A^{(1,0)} + A^{(0,1)}$ 
where $A^{(1,0)*} = -A^{(0,1)}$. 
Thus we can identify the space of unitary connections with its $(0,1)$ part 
and the tangent space is also $(0,1)$ part of a $1$-form
Let $\alpha^{(0,1)}, \beta^{(0,1)} \in T_{A} {\mathcal A} = \Omega^{(0,1)} (M, {\rm ad} P\otimes \CC) $, such that $\alpha^{(1,0)} = - \alpha^{(0,1)*}$ and   $\beta^{(1,0)} = -\beta^{(0,1)*}$. 
Let
$\gamma^{(1,0)}, \delta^{(1,0)} \in T_{\Phi} {\mathcal H} = \Omega^{(1,0)} (M, \rm{ad} P \otimes \CC )$.
Let us extend $\gamma^{(1,0)}, \delta^{(1,0)}$ by defining 
$\gamma^{(0,1)} = -\gamma^{(1,0)*}$ and $\delta^{(0,1)} = - \delta^{(1,0)*}$. 
Thus, $\alpha = \alpha^{(0,1)} + \alpha^{(1,0)}$ and $ \beta = \beta^{(0,1)} +
\beta^{(1,0)}$, $\gamma = \gamma^{(0,1)} + \gamma ^{(1,0)}$ and $\delta = \delta^{(0,1)} + \delta^{(1,0)}$, i.e.  $\alpha, \beta, \gamma, \delta  \in \Omega^1(M, {\rm ad} P).$
Let $X, Y$ be two tangent vectors to the configuration space, given
 by $X= (\alpha^{(0,1)}, \gamma^{(1,0)})$, and $Y= (\beta^{(0,1)} , \delta^{(1,0)})$.
As in ~\cite{D}, let us define a metric on the complex configuration space
\begin{eqnarray*}
& &  g (X, Y) = g ((\alpha^{(0,1)}, \gamma^{(1,0)}), (\beta^{(0,1)} , \delta^{(1,0)}))  \\  &=&  -   \int_M {\rm Tr} (   \alpha \wedge *_1 \beta) - 2 {\rm Im} 
\int_M {\rm Tr}  ( \gamma^{(1,0)} \wedge *_2 \delta^{(1,0){\rm tr}}) \\
&=& -2{\rm Im}  \int_M {\rm Tr} (\alpha^{(0,1)} \wedge \beta^{(1,0)}) - 2 {\rm Im}  \int_M {\rm Tr} (\gamma^{(1,0)} \wedge \delta^{(1,0)*}) \\
&=& 2{\rm Im}  \int_M {\rm Tr} (\alpha^{(0,1)} \wedge \beta^{(0,1)*}) - 2 {\rm Im}  \int_M {\rm Tr} (\gamma^{(1,0)} \wedge \delta^{(1,0)*}) {\rm \;\;\;\;\;\;\;\;\;\;\;\;\;\;\;\;\;\;\;\;\;\;\;\;\;\;(3)}
\end{eqnarray*}
 Here the superscript ${\rm tr} $ 
stands for  tranpose in the Lie algebra of $U(n)$, $*_1$ denotes 
the Hodge star taking $dx$ forms to $dy$ forms and $dy$ forms to $-dx$ forms 
(i.e. $*_1 (\eta dz) = -i \eta dz$ and 
$*_1(\bar{\eta} d \bar{z}) = i \bar{\eta} d \bar{z}$)  and 
 $*_2$ denotes the  operation (another Hodge star), 
 such that $*_2(\eta dz) = \bar{\eta} d \bar{z}$ and 
$*_2(\bar{\eta} d \bar{z}) =  -\eta d z$.
To get the metric in its final form, we have used the fact that 
\begin{eqnarray*}
& & 2 {\rm Im} \int_M {\rm Tr} (\alpha^{(0,1)} \wedge \beta^{(1,0)})\\
&=& \frac{1}{i}  \int_M {\rm Tr} [\alpha^{(0,1)} \wedge \beta^{(1,0)} - 
 \overline{\alpha^{(0,1)}} \wedge \overline{\beta^{(1,0)}}]\\  
&=& \frac{1}{i} \int_M {\rm Tr} [\alpha^{(0,1)} \wedge \beta^{(1,0)} -  (-\overline{\alpha^{(0,1)\rm tr}}) \wedge (-\overline{\beta^{(1,0)\rm tr}})] \\
&=& -i \int_M {\rm Tr} [\alpha^{(0,1)} \wedge \beta^{(1,0)} -  \alpha^{(1,0)} \wedge \beta^{(0,1)}] \\
&=& \int_M {\rm Tr} (\alpha \wedge *_1 \beta )
\end{eqnarray*} 
Note that the metric can also be expanded and written in the form
\begin{eqnarray*}
  g (X,Y) &=& i [ \int_M {\rm Tr} (\alpha^{0,1} \wedge \beta^{1,0} -  \alpha^{1,0} \wedge \beta^{0,1})  \\
&-& \int_M {\rm Tr} (\gamma^{1,0} \wedge \delta ^{0,1} +  \gamma^{0,1} \wedge \delta^{1,0} ) ]  
\end{eqnarray*}

We check that this coincides with the  metric on the moduli space
${\mathcal M}$ given by ~\cite{H}, page 79 and page 88. 
On $ T_{(A, \Phi)} {\mathcal C} = T_A {\mathcal A} \times T_{\Phi} {\mathcal H}$which is $\Omega^{(0,1)}(M, {\rm ad} P \otimes \CC) \times 
\Omega^{(1,0)}(M, {\rm ad} P \otimes \CC)$, Hitchin defines 
a metric $g_1$ such that 

$g_1((\alpha^{(0,1)},\gamma^{(1,0)}), (\alpha^{(0,1)}, \gamma^{(1,0)})) 
= 2i  \int_M {\rm Tr} (\alpha^{(0,1)*} \wedge \alpha^{(0,1)})
+ 2i  \int_M {\rm Tr} (\gamma^{(1,0)} \wedge \gamma^{(1,0)*})$.   

$*$ denotes conjugate transpose as usual. 
Let $ \gamma^{(1,0)}= c dz$, where $c$ is a matrix. 
On $T_{(A, \Phi)} {\mathcal C}$, our metric 
\begin{eqnarray*}
& & g((\alpha, \gamma^{(1,0)}), (\alpha, \gamma^{(1,0)})) \\ 
&=&  -  \int_M   {\rm Tr} (\alpha \wedge *_1 \alpha) 
- 2 {\rm Im}  \int_M {\rm Tr} ( \gamma^{(1,0)} \wedge *_2 \gamma^{(1,0){\rm tr}})\\
&=& -  \int_M {\rm Tr} [(\alpha^{(1,0)} + \alpha^{(0,1)}) 
\wedge (-i \alpha^{(1,0)} + i \alpha^{(0,1)})] \\
& & -2 {\rm Im}   \int_M {\rm Tr} (c dz \wedge c^* d \bar{z}) \\
&=& 2i  \int_M {\rm Tr} (\alpha^{(0,1)} \wedge \alpha^{(1,0)}) 
-2 Im    \int_M (-2i){\rm Tr} (c c^*) dx \wedge dy  \\
&=& -2i  \int_M {\rm Tr} (\alpha^{(0,1)} \wedge \alpha^{(0,1)*}) 
+ 4 \int_M Re ({\rm Tr} (c c^*)) dx \wedge dy  \\
&=& 2i  \int_M {\rm Tr} (\alpha^{(0,1)*} \wedge \alpha^{(0,1)}) 
+ 2i \int_M (-2i) {\rm Tr} (c c^*) .  dx \wedge dy \\
 &=& 2i  \int_M {\rm Tr} (\alpha^{(0,1)*} \wedge \alpha^{(0,1)}) 
+ 2i \int_M {\rm Tr} (c c^*)  dz \wedge d \bar{z} \\
&=& 2i \int_M {\rm Tr} (\alpha^{(0,1)*} \wedge \alpha^{(0,1)})
+ 2i  \int_M {\rm Tr} (\gamma^{(1,0)} \wedge \gamma^{(1,0)*}) \\
\end{eqnarray*} 
where we have used the fact that $\alpha^{(1,0)} = -\alpha^{(0,1)*}$ and that 
${\rm Tr} (cc^*)$ is real.  
Thus we get the same metric as Hitchin does.

To give the complex structures of the hyperK\"{a}hler structure explicitly, 
(which is missing in ~\cite{H}), let us define three almost complex structures 
acting on the tangent space to the 
configuration space, i.e. acting on $T = \Omega^{(0,1)} (M, {\rm ad} P \otimes \CC) \times \Omega^{(1,0)}(M, {\rm ad} P \otimes \CC )$, 
$${\mathcal I} = \left[ \begin{array}{cc}
i   & 0 \\
 0  & i
\end{array} \right], $$ 
$${\mathcal J} = \left[ \begin{array}{cc}
0  & i \tilde{*}_2 \\
 i \tilde{*}_2 & 0
\end{array} \right], $$ 
$${\mathcal K} = \left[ \begin{array}{cc}
0  & -\tilde{*}_2 \\
 - \tilde{*}_2 & 0
\end{array} \right] $$ 
where $\tilde{*}_2 (\alpha)$  $=$ $*_2 \alpha^{\rm tr}$ 
is such that $\tilde{*}_2 (i\alpha)$ $ =$ $- i\tilde{*}_2\alpha$. 

{\bf Note:} A more detailed description of the three complex structures 
is given in the  section $(4)$ where we discuss holomorphicity of the 
prequantum line bundles. There we give it in physicists' 
notation commensurate with ~\cite{KW}.
  
These three complex structures satisfy the 
quarternionic algebra of matrices acting on 
$T = \Omega^{(0,1)} (M, {\rm ad} P \otimes \CC) \times \Omega^{(1,0)}(M, {\rm ad} P \otimes \CC)$:
$$ {\mathcal I}^2 = {\mathcal J}^2 = {\mathcal K}^2 = -1, $$
$$ {\mathcal I} {\mathcal J} = - {\mathcal J}{\mathcal I} = {\mathcal K},$$
$$ {\mathcal J}{\mathcal K} = - {\mathcal K}{\mathcal J} = I, $$
$${\mathcal K}{\mathcal I} = - {\mathcal I}{\mathcal K} = {\mathcal J}.$$
We define three symplectic forms as follows:
$$\Omega (X,Y) = g (X, {\mathcal I}Y),$$
$$ {\mathcal Q}_1 (X, Y) = g (X, {\mathcal J}Y),$$ 
$$ {\mathcal Q}_3 (X, Y) = g (X, {\mathcal K}Y) $$

Let $X = (\alpha^{(0,1)}, \gamma^{(1,0)})$, $Y=(\beta^{(0,1)}, \delta^{(1,0)})$
be two tangent vectors belonging to $T$.
\begin{eqnarray*}
& & \Omega ((\alpha^{(0,1)}, \gamma^{(1,0)}), (\beta^{(0,1)}, \delta^{(1,0)})) 
= g((\alpha^{(0,1)}, \gamma^{(1,0)}), (i\beta^{(0,1)}, i\delta^{(1,0)})) \\
&=& 2{\rm Im}  \int_M {\rm Tr} (\alpha^{(0,1)} \wedge (i \beta^{(0,1)})^*)
- 2 {\rm Im}  \int_M {\rm Tr} (\gamma^{(1,0)} \wedge (i\delta^{(1,0)})^* )\\
&=& 2 {\rm Re}  \int_M {\rm Tr} (\alpha^{(0,1)} \wedge \beta^{(1,0)}) -
2 {\rm Im}  \int_M {\rm Tr} ((i) \gamma^{(0,1)} \wedge (\delta^{(0,1)})\\
&=& \int_M {\rm Tr} (\alpha \wedge \beta)
- 2 {\rm Re}  \int_M {\rm Tr} (\gamma^{(1,0)} \wedge \delta^{(0,1)}) \\
&=&  \int_M {\rm Tr} (\alpha \wedge \beta) -  \int_M {\rm Tr} (\gamma \wedge \delta)
\end{eqnarray*}
where we have used the fact that 
$$ 2 {\rm Re}  \int_M {\rm Tr} (\alpha^{(0,1)} \wedge \beta^{(1,0)}) =  
\int_M {\rm Tr} (\alpha \wedge \beta) $$
which follows from the fact that 
\begin{eqnarray*}
 \int_M {\rm Tr} (\overline{\alpha^{(0,1)}} \wedge \overline{\beta^{(1,0)}})
&=&  \int_M {\rm Tr} [(-\overline{\alpha^{(0,1)\rm tr}}) \wedge (-\overline{\beta^{(1,0)\rm tr}})] \\
&=&  \int_M {\rm Tr} (\alpha^{(1,0)} \wedge \beta^{(0,1)}) {\rm \;\;\;\;\;\;\;\;\;\;\;\;\;\;\;\;\;\;\;\;\;\;\;\;\;\;\;\;\;\;\;\;\;\;\;\;\;\;\;\;\;\;\;\;\;\;\;(4)}
\end{eqnarray*}

Following the ideas in ~\cite{H}, we had shown in ~\cite{D}  by a moment map 
construction that this form descends to a  symplectic form on the moduli 
space ${\mathcal M}$. (The explicit construction of this form is missing in 
~\cite{H}). The first equation, i.e. eqn $(1)$ gives the moment map for this
symplectic form.

\begin{eqnarray*}
{\mathcal Q}_1 (X, Y)  &=& g(X, {\mathcal J} Y) = g ((\alpha^{(0,1)}, \gamma^{(1,0)}), (i \tilde{*}_2 \delta^{(1,0)},
i \tilde{*}_2 \beta^{(0,1)})) \\
&=& g ((\alpha^{(0,1)}, \gamma^{(1,0)}), (-i  \delta^{(0,1)},
i  \beta^{(1,0)})) \\
&=& -2 {\rm Re} \int_M {\rm Tr} (\alpha^{(0,1)} \wedge \delta^{(1,0)})
-2 {\rm Re} \int_M {\rm Tr} (\gamma^{(1,0)} \wedge \beta^{(0,1)}) \\
&=& - [  \int_M {\rm Tr} (\alpha \wedge \delta) +  \int_M {\rm Tr} (\gamma \wedge \beta)]
\end{eqnarray*}
\begin{eqnarray*}
 {\mathcal Q}_2 (X, Y) &=& g(X, {\mathcal K} Y) = g ((\alpha^{(0,1)}, \gamma^{(1,0)}), (- \tilde{*}_2 \delta^{(1,0)},
- \tilde{*}_2 \beta^{(0,1)})) \\
&=& g ((\alpha^{(0,1)}, \gamma^{(1,0)}), (  \delta^{(0,1)},
- \beta^{(1,0)})) \\
&=&-2 {\rm Im}  \int_M {\rm Tr} (\alpha^{(0,1)} \wedge \delta^{(1,0)})
-2 {\rm Im}  \int_M {\rm Tr} (\gamma^{(1,0)} \wedge \beta^{(0,1)})\\
&=& \int_M {\rm Tr} (\alpha \wedge \tilde{\delta} + \tilde{\gamma} \wedge \beta)\end{eqnarray*}
where $\tilde{\delta} = i (\delta^{(1,0)} - \delta^{(0,1)})$ and
$\tilde{\gamma} = i (\gamma^{(1,0)} - \gamma^{(0,1)}).$
Next, following ~\cite{H}, page 90, we define a symplectic form 
\begin{eqnarray*}
{\mathcal Q} (X,Y) &=& 2 {\rm Tr} \int_M (
\delta^{(1,0)} \wedge \alpha^{(0,1)} - \gamma^{(1,0)} \wedge   \beta^{(0,1)})\\
&=& ({\mathcal Q}_1 + i {\mathcal Q}_2) (X,Y)
\end{eqnarray*}

In ~\cite{H}, Hitchin shows that this form $ {\mathcal Q}$ 
descends as a symplectic form  to the moduli space of solution of the 
self-duality equations. He proves this using a moment map construction, i.e.
the second equation,  eqn $(2)$, gives the moment map for this symplectic 
form. (Note that the factor of two doesnot alter the moment map construction).
Using his method, it can be shown  that the metric $g$ descends to the 
moduli space ${\mathcal M}$ and is hyperK\"{a}hler and the three symplectic 
forms are exactly $\Omega, {\mathcal Q}_1, {\mathcal Q}_2$.

{\bf  Exactness of forms ${\mathcal Q}_1$ and ${\mathcal Q}_2$:}

In this section we first establish a link between the notation in ~\cite{KW} 
(page 40-44) and our notation. 
Will take the example of ${\mathcal Q}_1$ and ${\mathcal Q}_2$ and explain the 
notation.

In the Kapustin Witten notation, $\Phi = \phi_z dz + \phi_{\bar{z}} d \bar{z} = \Phi^{1,0} + \Phi^{0,1}$ and let $X = (\alpha, \gamma)$, $Y = (\beta, \delta).$
We will denote by $\omega_1$ their $\omega_K$ and by $\omega_2$ their 
$\omega_J$. Their $J$ is our $-K$ and their $K$ is our $J$, their $\omega_I$ is $\frac{-1}{2 \pi} \Omega$ and we will just now show that their $\omega_J$ is $\frac{1}{2 \pi} Q_2$ and $\omega_K$ is $\frac{-1}{2 \pi} Q_1$. 
\begin{eqnarray*}
\omega_1 (X, Y) &=& \frac{i}{2\pi}   \int_M  i
(dz \wedge d \bar{z}) {\rm Tr} [(\delta \phi_{\bar{z}} \otimes \delta A_{z} - 
\delta A_z \otimes \delta \phi_{\bar{z}}) \\
&-&  (\delta \phi_z \otimes \delta A_{\bar{z}} - \delta A_{\bar{z}} \otimes \delta \phi_z)] (X, Y)\\
&=& \frac{-1}{2 \pi}  [\int_M {\rm Tr} (-\gamma^{0,1} \wedge \beta^{1,0} - \alpha^{1,0} \wedge \delta^{0,1}) \\
&-& {\rm Tr} (\gamma^{1,0} \wedge \beta^{0,1} + \alpha^{0,1} \wedge \delta^{1,0})] \\
&=& \frac{1}{2 \pi} \int_M \rm{Tr} (\gamma \wedge \beta + \alpha \wedge \delta)\\
&=&\frac{-1}{2 \pi} {\mathcal Q}_1(X, Y)
\end{eqnarray*}

Similarly,
\begin{eqnarray*}
\omega_2 (X, Y) &=& \frac{1}{2 \pi} \int_M i (dz \wedge d \bar{z}) {\rm Tr} [ (\delta \phi_{\bar{z}} \wedge \delta A_z) \\
&+& (\delta \phi_z \wedge \delta A_{\bar{z}})] (X, Y) \\
&=& \frac{1}{2 \pi}  \int_M i (dz \wedge d \bar{z}) {\rm Tr} [(\delta \phi_{\bar{z}} \otimes \delta A_z - \delta A_z \otimes \delta \phi_{\bar{z}}) \\
&+& (\delta \phi_z \otimes \delta A_{\bar{z}} - \delta A_{\bar{z}} \otimes \delta \phi_z )] (X, Y)\\
&=& \frac{i}{2 \pi}  \int_M {\rm Tr} [(- \gamma^{0,1} \wedge \beta^{1,0} - \alpha^{1,0} \wedge \delta^{0,1}) \\
&+&  \gamma^{1,0} \wedge \beta^{0,1} + \alpha^{0,1} \wedge \delta^{1,0})]\\
&=& \frac{1}{2 \pi} {\mathcal Q}_2 (X, Y)
\end{eqnarray*}

Now as in  ~\cite{KW}, page 44,
${\mathcal Q}_1 = \delta \theta_1 $ where 
\begin{eqnarray*}
\theta_1 &=&  -i \int_M i (dz \wedge d \bar{z}) {\rm Tr} ( \phi_{\bar{z}} \delta A_z - \phi_z \delta A_{\bar{z}}) \\
&=&   -\int_M {\rm Tr} (\Phi \wedge \delta A) 
\end{eqnarray*}
s.t.
$\theta_1 (\alpha, \gamma) =  -\int_M {\rm Tr} ( \Phi \wedge \alpha ) $

Similarly, ${\mathcal Q}_2 = \delta \theta_2$ where 
$\theta_2 =  \int_M i (dz \wedge d \bar{z}) {\rm Tr} ( \phi_{\bar{z}} \delta A_z + \phi_z \delta A_{\bar{z}}) $ such that  
$\theta_2 (\alpha, \gamma) =  \int_M {\rm Tr} (\Phi^{1,0} \wedge \alpha^{0,1} - \Phi^{0,1} \wedge \alpha^{1,0}). $
 
Now $\theta_1$ and $\theta_2$ both descend as $1$-forms on the moduli space ${\mathcal M}$ since they are gauge-invariant. Recall  $\Phi$, $\tilde{\Phi}$,
$\alpha$  transform by the adjoint 
representation of  $U(n)$ (unlike $A$) and keeps the $1$-forms $\theta_1$ and $\theta_2$ gauge invariant (since we are taking trace).

Thus ${\mathcal Q}_{1}$ and ${\mathcal Q}_{2}$ are both exact.

\section{Prequantum line bundles}

In this section we review Quillen's determinant line bundle of Cauchy-Riemann 
operators which will enable us to construct prequantum line bundles 
corresponding to the hyperK\"{a}hler structure in the Hitchin system. 

First let us note that a connection $A$ on a principal bundle induces a  
connection on any associated vector bundle $E$. 
We will denote this connection also by $A$ since  the same ``
Lie-algebra valued $1$-form'' $A$ (modulo representations)  gives  a covariant 
derivative operator enabling you to take derivatives of  sections of $E$ 
~\cite{N}, page 348.
A very clear description of
the determinant line bundle can be found in ~\cite{BF} and
~\cite{Q}. Here we mention  the formula for the Quillen curvature
of the determinant line bundle $\wedge^{\rm {top}}({\rm Ker} \bar{\partial}_A)^{*}
\otimes \wedge^{{\rm top}}({\rm Coker} \bar{\partial}_A) = 
{\rm det}(\bar{\partial}_A),
$ where $\bar{\partial}_A = \bar{\partial} + A^{(0,1)}$ , 
given the canonical unitary
connection $\nabla_Q$, induced by the Quillen metric~\cite{Q}.
Recall that the affine space ${\mathcal A}$ (notation as
in ~\cite{Q}) is an infinite-dimensional K\"{a}hler manifold. Here
each connection  is identified with its $(0,1)$ part. Since the
total connection is unitary (i.e. of the form $A= A^{(1,0)} + A^{(0,1)}$, 
where $A^{(1,0)} = -A^{(0,1)*}$) this identification is easy.
 In fact, for every $A \in {\mathcal A}$,
$T_A^{\prime} ({\mathcal A}) = \Omega^{0,1} (M, {\rm ad} P \otimes \CC)$ and 
the corresponding K\"{a}hler form is given by 
$$F(\alpha, \beta) = 2 {\rm Re}  \int_M  {\rm Tr} (\alpha^{(0,1)} \wedge  
\beta^{(0,1)*}) = - 2{\rm Re} \int_M  {\rm Tr} (\alpha^{(0,1)} \wedge  
\beta^{(1,0)}) $$   
where $\alpha^{(0,1)}, \beta^{(0,1)} \in \Omega^{0,1} (M, {\rm ad} P \otimes \CC)$, and $\beta^{(1,0)} = - \beta^{(0,1)*}.$     It is skew symmetric 
if you interchange 
$\alpha^{(0,1)} = A d \bar{z}$ and $\beta^{(0,1)} = B d \bar{z}$ 
(follows from the fact that 
${\rm Im} ({\rm Tr} (A B^*)) = - {\rm Im} ({\rm Tr} (B A^*))$ for matrices $A$ 
and $B$, using once again $ d \bar{z} \wedge d z $ is imaginary).  
Let $\alpha = \alpha^{(0,1)} + \alpha^{(1,0)}$, $\beta =\beta^{(0,1)} + 
\beta^{(1,0)}$. 
It is clear from the fact that $ \alpha^{(1,0)} = -\alpha^{(0,1)*} $ and 
$\beta^{(1,0)} = -\beta^{(0,1)*}$ we have $$ F(\alpha, \beta) = - \int_M {\rm Tr} (\alpha \wedge \beta). $$
(see for instance, ~\cite{H1}, page 358). 
Let $\nabla_Q$ be the connection induced from the Quillen metric and 
${\mathcal F}(\nabla_Q)$ be the Quillen curvature. Then one has,
$$ {\mathcal F}(\nabla_Q) =  \frac{i}{ 2\pi}  F $$ 

\subsection{Quantization of the moduli space ${\mathcal M}$}
In this section, we will show that for each of the symplectic forms 
$\Omega$, ${\mathcal Q}_1$ and ${\mathcal Q}_2$ there are prequantum line 
bundles ${\mathcal P}$, ${\mathcal E}$, ${\mathcal N}$ whose respective 
curvatures are  these symplectic forms.

First we note that to the connection $A$ we can add any one form and
still obtain a derivative operator. 

On the principal bundle $P$ on the Riemann surface, 
 we can define new connections,  $A \pm \Phi = A^{(0,1)} + A^{(1,0)} \pm \Phi^{(0,1)} \pm \Phi^{(1,0)}$ 
 and $A \mp \tilde{\Phi} = A^{(0,1)} + A^{(1,0)} \pm i \Phi^{(0,1)} \mp i \Phi^{(1,0)}$ where $\tilde{\Phi} = i (\Phi^{(1,0)} - \Phi^{(0,1)})$, i.e. 
$\tilde{\Phi}^{(0,1)} = -i \Phi^{(0,1)}$ and 
$ \tilde{\Phi}^{(1,0)} = i \Phi^{(1,0)}$. (Note that, as usual, 
$\Phi^{(1,0)*} = - \Phi^{(0,1)}$ and $\tilde{\Phi}^{(1,0)*} = - \tilde{\Phi}^{(0,1)}$).

{\bf Definitions} 

 Let us denote by ${\mathcal L} = 
{\rm det} (\bar{\partial}+ A^{0,1})$ a determinant bundle on ${\mathcal A}$. 

Let  ${\mathcal R} = {\rm det} (\bar{\partial} + A_0^{(0,1)} 
+ \Phi^{(0,1)})$  
where $A_0$ is a connection whose gauge equivalence class is fixed, 
i.e. $A_0$ is allowed to change only in the gauge direction. 

Let ${\mathcal P} = {\mathcal L}^{-2} \otimes  
 {\mathcal R}^2$ denote a line bundle 
over ${\mathcal C} = {\mathcal A} \times {\mathcal H}$. 

(This combination will give the prequantum line bundle corresponding to 
$\Omega$). 
Let us define  
${\mathcal E}_{\pm} = {\rm det}(\bar{\partial} + A^{(0,1)} \pm \Phi^{(0,1)})$ on  the affine space 
${\mathcal B}_{\pm} = \{ A^{(0,1)} \pm \Phi^{(0,1)} | 
A^{(0,1)} \in {\mathcal A},  \Phi^{(0,1)*} = - \Phi^{(1,0)} \in {\mathcal H} \}$ 
which is 
isomorphic to 
${\mathcal C} = \{ A^{(0,1)} \in {\mathcal A}\} \times \{ 
\Phi^{(1,0)} \in {\mathcal H} \}  = {\mathcal A} \times {\mathcal H}.$

Let ${\mathcal E} = {\mathcal E}_{+} \otimes ({\mathcal E}_{-})^{-1}$
(We take this  combination because it will 
 give the prequantum line corresponding to ${\mathcal Q}_1$.)

Similarly let us define 
${\mathcal N}_{\pm} =  {\rm det}(\bar{\partial} + A^{(0,1)} \pm i\Phi^{(0,1)})$ on  
the affine space 

${\mathcal V}_{\pm} = \{ A^{(0,1)} \pm i\Phi^{(0,1)} | 
A^{(0,1)} \in {\mathcal A},  \Phi^{(0,1)*}=-\Phi^{(1,0)} \in {\mathcal H} \}$ 
which is 
isomorphic to ${\mathcal C} = {\mathcal A} \times {\mathcal H}$. 

Let ${\mathcal N} = {\mathcal N}_{+} \otimes ({\mathcal N}_{-})^{-1}$ 

(Once, again this will be the prequantum line bundle corresponding
to ${\mathcal Q}_2$).

\begin{lemma}
${\mathcal P}$, ${\mathcal E}_{\pm}$ and ${\mathcal N}_{\pm}$ are  well-defined line bundles 
over ${\mathcal M} \subset {\mathcal C}/{\mathcal G}$, 
where ${\mathcal G}$ is the gauge group.
\end{lemma}
\begin{proof}
Let us consider the Cauchy-Riemann operator 
$ D= \bar{\partial} + A^{(0,1)} + \Phi^{(0,1)}$ which appears in 
${\mathcal E}_{+}$. All the other cases are analogous.
 Under gauge transformation 
$D=\bar{\partial} + A^{(0,1)} + \Phi^{(0,1)}  
\rightarrow D_g= g(\bar{\partial} + A^{(0,1)} + \Phi^{(0,1)})g^{-1} $ since it is the $(0,1)$ part of the connection operator $d + A + \Phi$ which 
transforms in the same way.
We can show that the operators $D$ and $D_g$ have isomorphic 
kernel and cokernel and their corresponding Laplacians have the 
same spectrum and the eigenspaces are of the same dimension. Let 
$\Delta$ denote the Laplacian corresponding to $D$ and $\Delta_g$ 
that corresponding to $D_g$.The Laplacian is $\Delta = \tilde{D} D$ where 
$\tilde{D} = \partial + A^{(1,0)} + \Phi^{(1,0)}$, where recall $A^{(1,0)*} = -A^{(0,1)}$ and $\Phi^{(1,0)*} = - \Phi^{(0,1)}$. Note that  $\tilde{D} \rightarrow  \tilde{D}_g = g \tilde{D} g^{-1}$ under gauge transformation since it is 
the $(1,0)$ part of the connection operator $ d+ A+ \Phi$ which transforms in 
the same way.
 Thus $\Delta_g = g \Delta g^{-1}$.  
Thus the isomorphism of eigenspaces of $\Delta$ and $\Delta_g$ 
is  $s \rightarrow g s$. We describe here how to define the line bundle on the 
moduli space. Let $K^a(\Delta)$ is the direct sum of 
eigenspaces of the operator $\Delta$ of 
eigenvalues $< a$, over the open subset 
$U^a = \{A^{(0,1)} + \Phi^{(0,1)} | a \notin {\rm Spec} \Delta \}$ of the affine space 
${\mathcal B}_{+}.$ The determinant line bundle is defined using the exact sequence
$$ 0 \rightarrow {\rm Ker} D \rightarrow K^a(\Delta) \rightarrow 
D(K^a(\Delta)) \rightarrow {\rm Coker} D \rightarrow 0$$ 
Thus 
one identifies 
${\rm det} D$= $\wedge^{{\rm top} }({\rm Ker} D)^* \otimes \wedge^{{\rm top} }
({\rm Coker} D)$ with 
 $\wedge^{{\rm top}}(K^a(\Delta))^* \otimes \wedge^{{\rm top}} 
(D(K^a(\Delta)))$ (see ~\cite{BF} for more details)
and there is an isomorphism of the fibers as $D \rightarrow D_g$. 
Thus one can identify 

$ \wedge^{{\rm top}}(K^a(\Delta))^* \otimes \wedge^{{\rm top}} 
(D(K^{a}(\Delta))) \equiv
\wedge^{{\rm top}}(K^a(\Delta_g))^* \otimes \wedge^{{\rm top}} 
(D(K^{a}(\Delta_g))).$

By extending this definition from 
$U^a$ to $V^a = \{(A^{(0,1)}, \Phi^{(1,0)})| a \notin {\rm Spec} \Delta \}$, 
an open subset of ${\mathcal C}$,  we can define the fiber over 
the quotient space $V^a/{\mathcal G}$ to be the 
equivalence class of this fiber. Covering ${\mathcal C}$ by open sets of the 
type $V^a$ enables us to define it on ${\mathcal C}/{\mathcal G}$. Then we 
restrict it to the moduli space ${\mathcal M} \subset 
{\mathcal C}/{\mathcal G}$.

Similarly one can deal with the other terms  in ${\mathcal E}$,  
${\mathcal P}$ and ${\mathcal N}$.
\end{proof}

{\bf Curvatures and symplectic forms}

Recall $\alpha \in \Omega ^{1}(M, {\rm ad} P)$ has the decomposition 
$\alpha = \alpha^{(1,0)} + \alpha^{(0,1)}$, where 
$\alpha^{(1,0)}= - \alpha^{(0,1)*}$ . Similar decomposition holds 
for $\beta, \gamma, \delta \in \Omega^{1}(M, {\rm ad} P)$.  

Let $p = (A, \Phi) \in S$ where $S$ is the space of solutions to Hitchin 
equations $(1)$ and 
$(2)$. Let $X, Y \in T_{[p]}{\mathcal M}$. We write $X =(\alpha, \gamma)$ and $Y=(\beta, \delta)$, where 
$\alpha^{(0,1)}, \beta^{(0,1)} \in T_{A} ({\mathcal A}^{(0,1)}) = \Omega^{(0,1)}(M, {\rm ad} P \otimes \CC)$ and 
$\gamma^{(1,0)}, \delta^{(1,0)} \in T_{\Phi} {\mathcal H} = \Omega^{(1,0)}(M, {\rm ad} P \otimes \CC)$.
Since $T_{[p]}{\mathcal M}$ can be identified with a subspace in 
$T_p S$ orthogonal to $T_p O_p$ (the tangent space to the gauge orbit)  
 then 
$X,Y$ can be said to satisfy (a) $X, Y \in T_p S$ i.e. they satisfy 
linearization of $(1)$ and $(2)$ and (b) $X, Y$ are orthogonal to $T_p O_p $, the tangent space to the gauge orbit.

Let ${\mathcal F}_{{\mathcal L}^{-2}}$, $ {\mathcal F}_{{\mathcal R}^2}$,  
 denote the Quillen curvatures of the
determinant line bundles ${\mathcal L}^{-2}$, ${\mathcal R}^2$,  
 respectively.
Then, by the Quillen formula in the previous section,
\begin{eqnarray*}
{\mathcal F}_{{\mathcal L}^{-2}}((\alpha, \gamma), (\beta, \delta)) 
&=& -2 {\mathcal F}_{{\mathcal L}}((\alpha, \gamma), (\beta, \delta))\\ 
&=& - 2 \frac{i}{\pi} {\rm Re} {\rm Tr}\int_M (\alpha^{(0,1)}  
\wedge \beta^{(0,1)*}) \\
&=& \frac{i}{\pi} {\rm Tr} \int_M \alpha \wedge \beta
\end{eqnarray*}

(Since there is no $\Phi$-term in ${\mathcal L}$, 
$\gamma$ and $\delta$ donot contribute).

\begin{eqnarray*}
{\mathcal F}_{{\mathcal R}^2}((\alpha, \gamma), 
(\beta, \delta)) 
&=& 2{\mathcal F}_{{\mathcal R}}((\alpha, \gamma), 
(\beta, \delta)) \\
&=& 2\frac{i}{\pi} {\rm Re}  \int_M  {\rm Tr} (\gamma^{(0,1)} \wedge \delta^{(0,1)*})\\
&=& -2\frac{i}{\pi} {\rm Re}  \int_M {\rm Tr} (\gamma^{(0,1)} \wedge \delta^{(1,0)})\\
&=& -2\frac{i}{\pi} {\rm Re}  \int_M  {\rm Tr} (\overline{(-\gamma^{(0,1)tr})} \wedge \overline{(-\delta^{(1,0)tr)})})\\
&=&-2\frac{i}{\pi} {\rm Re}  \int_M {\rm Tr} ( \gamma^{(1,0)} \wedge \delta^{(0,1)})\\
&=& 2\frac{i}{\pi} {\rm Re}  \int_M {\rm Tr} (\gamma^{(1,0)} \wedge 
\delta^{(1,0)*} )\\
&=& -\frac{i}{\pi}   \int_M  {\rm Tr} (\gamma \wedge \delta)
\end{eqnarray*}

Note:  $\gamma^{(0,1)}$ and $\delta^{(0,1)*}$ contributes because  of the term $\Phi^{(0,1)}$ in the C-R operator in ${\mathcal R}$. $\alpha$, $\beta$  donot contribute to this curvature  
because in the definition of ${\mathcal R}$ the gauge equivalence 
class of $A_0$ is fixed. 

It is easy to check that the curvature of ${\mathcal P}$ is 
 $${\mathcal F}_{{\mathcal L}^{-2}}  
+ {\mathcal F}_{{\mathcal R}^2} 
= \frac{i}{\pi} \Omega.$$

The line bundles ${\mathcal E}_{\pm}$  are determinant of C-R operators of 
connections  $A^{(0,1)} \pm \Phi^{(0,1)}$. Hence, by the formula is the previous section,  in the Quillen curvature of ${\mathcal E}{\pm}$ terms like 
$\alpha \pm \gamma$ and $ \beta \pm \delta$ will appear. 

The Quillen curvature of ${\mathcal E}_{\pm}$ is 
\begin{eqnarray*}
& &{\mathcal F}_{{\mathcal E}_{\pm}}((\alpha, \gamma), (\beta, \delta))  = \frac{-i}{2\pi} ( \int_M {\rm Tr} [ (\alpha \pm \gamma) \wedge (\beta \pm \delta)]) \\
&=&\frac{-i}{2\pi}(   \int_M {\rm Tr} [\alpha \wedge \beta \pm   \gamma \wedge 
\beta \pm  \alpha \wedge \delta +  \gamma \wedge \delta])
\end{eqnarray*}

Thus curvature of ${\mathcal E} = {\mathcal E}_{+} \otimes ({\mathcal E}_{-})^{-1}$ is 
\begin{eqnarray*}
({\mathcal F}_{{\mathcal E}_{+}} - {\mathcal F}_{{\mathcal E}_{-}})( (\alpha, \gamma), (\beta, \delta)) & =& 
 \frac{-i}{\pi}[ \int_M {\rm Tr} (\alpha \wedge \delta +  \gamma \wedge \beta) \\&=& \frac{i}{\pi}{\mathcal Q}_1((\alpha, \gamma), (\beta, \delta))\end{eqnarray*}
  Define $\tilde{\gamma} = 
i \gamma^{(1,0)} - i \gamma^{(0,1)}$. Similarly define $\tilde{\delta}$.

${\mathcal N}_{\pm}$ are determinant lines of Cauchy-Riemann operators of connections   $A^{(0,1)} \mp \tilde{\Phi}^{(0,1)}.$ Thus, by formula,  in the Quillen curvature, we will have 
terms like  $\alpha \mp \tilde{\gamma}$ and $ \beta \mp \tilde{\delta}$.
The Quillen curvature of ${\mathcal N}_{\pm}$ is 
\begin{eqnarray*}
& & {\mathcal F}_{{\mathcal N}_{\pm}}((\alpha, \gamma), (\beta, \delta))
= \frac{-i}{2\pi}(  \int_M {\rm Tr} [ (\alpha \mp \tilde{\gamma}) \wedge (\beta \mp \tilde{\delta})]) \\
&=& \frac{-i}{2\pi} \int_M {\rm Tr} [\alpha \wedge \beta \mp  \tilde{\gamma} \wedge 
\beta \mp  \alpha \wedge \tilde{\delta} +  \tilde{\gamma} \wedge \tilde{\delta}])
\end{eqnarray*}

The Quillen curvature of the line bundle ${\mathcal N} = {\mathcal N}_{+} \otimes ({\mathcal N}_{-})^{-1}$ is 
\begin{eqnarray*}
({\mathcal F}_{{\mathcal N}_{+}} - {\mathcal F}_{{\mathcal N}_{-}})((\alpha, \gamma), (\beta, \delta)) & =& 
\frac{i}{\pi}  \int_M {\rm Tr} (\alpha \wedge \tilde{\delta} +  \tilde{\gamma} \wedge \beta) \\
&=& \frac{i}{\pi}{\mathcal Q}_2 ((\alpha, \gamma), (\beta, \delta))
\end{eqnarray*}

Thus we have proved the following theorem.

\begin{theorem} The moduli space of solutions ${\mathcal M}$ admits three
prequantum  line bundles ${\mathcal P}$, ${\mathcal E}$ and ${\mathcal N}$ 
such that 
their Quillen curvatures are respectively the three sympletic forms, 
$\frac{i}{\pi}\Omega$, $\frac{i}{\pi}{\mathcal Q}_1$ and $\frac{i}{\pi}{\mathcal Q}_2$ which correspond to 
the hyperK\"{a}hler structure in ${\mathcal M}$.
\end{theorem} 

\section{Holomorphicity and Polarization} 

\begin{proposition}
${\mathcal P}^{-1}$ is a
${\mathcal I}$-holomorphic, ${\mathcal E}^{-1}$ is a ${\mathcal J}$-holomorphic 
and ${\mathcal N}^{-1}$ is a ${\mathcal K}$-holomorphic prequatum line bundle 
with curvature $-\frac{i}{\pi}\Omega$, $-\frac{i}{\pi}{\mathcal Q}_1$ and $-\frac{i}{\pi}{\mathcal Q}_2$ respectively.
\end{proposition}

\begin{proof}
Recall that $${\mathcal I} (\alpha^{(0,1)}) = i \alpha^{(0,1)},$$
$${\mathcal I}(\gamma^{(1,0)}) = i \gamma^{(1,0)},$$
$${\mathcal I}(\alpha^{(1,0)}) = -i \alpha^{(1,0)},$$
$${\mathcal I} (\gamma^{(0,1)}) = -i \gamma^{(0,1)}$$.

Thus w.r.t. ${\mathcal I},$  $A^{(0,1)}$ is holomorphic and $\Phi^{(0,1)}$ is antiholomorphic. Thus ${\mathcal L}$ is holomorphic and ${\mathcal R}$ is 
anti-holomorphic, by the same argument as in ~\cite{Q}.
But  ${\mathcal P}^{-1} = {\mathcal L}^2 \otimes {\mathcal R}^{-2}$ has the 
$A^{(0,1)}$-term as it is and the $\Phi^{(0,1)}$-term  in the inverse
bundle. Thus ${\mathcal P}^{-1}$ is ${\mathcal I}$-holomorphic.

Secondly, $${\mathcal J}(\alpha^{(0,1)}) = -i \gamma^{(0,1)},$$
 $${\mathcal J} (\gamma^{(1,0)}) = i \alpha^{(1,0)},$$
$$ {\mathcal J} (\alpha^{(1,0)}) = i \gamma^{(1,0)},$$
$${\mathcal J} (\gamma^{(0,1)}) = -i \alpha^{(0,1)}. $$

Thus w.r.t. ${\mathcal J},$ the $A^{(0,1)} - \Phi^{(0,1)}$-term 
is holomorphic and the $A^{(0,1)} + \Phi^{(0,1)}$-term is anti-holomorphic.
Thus ${\mathcal E}^{-1} = {\mathcal E}_{+}^{-1} \otimes {\mathcal E}_{-}$ is 
holomorphic since the anti-holomorphic term comes in the inverse.

Thirdly, $$ {\mathcal K} (\alpha^{(0,1)}) = \gamma^{(0,1)}, $$
$${\mathcal K} (\gamma^{(1,0)}) = - \alpha^{(1,0)}, $$
$${\mathcal K} (\alpha^{(1,0)}) = \gamma^{(1,0)},$$
$${\mathcal K} (\gamma^{(0,1)}) = - \alpha^{(0,1)}.$$ 

Thus w.r.t. ${\mathcal K}, $ the $A^{(0,1)} + i \Phi^{(0,1)}$-term is 
anti-holomorphic 
and the $A^{(0,1)} - i \Phi^{(0,1)}$-term is holomorphic.
Thus ${\mathcal N}^{-1} = {\mathcal N}^{-1}_{+} \otimes {\mathcal N}_{-}$ is 
holomorphic.
\end{proof}

{\bf Polarization:}

Since the symplectic forms are all K\"{a}hler,
we can take square integrable ${\mathcal I}$-holomorphic sections of 
${\mathcal P}^{-1}$ , ${\mathcal J}$-holomorphic sections of 
${\mathcal E}^{-1}$ and ${\mathcal K}$-holomorphic   sections of 
${\mathcal N}^{-1}$ as our Hilbert spaces. But we are still not guaranteed 
finite dimensional Hilbert spaces.

\section{Cherns-Simons gauge theory with complexified gauge group}

We introduce here the  Chern-Simons gauge theory with complexified gauge group
since flat connections on a principal bundle with complexified gauge group are
essentially solutions of self-duality equations,  as we shall 
elaborate below. In ~\cite{Wi1}, Witten had explored this -- quantizing one of 
the symplectic forms. 
By introducing a parameter $\lambda$ we wish to get all three 
the symplectic forms 
and all the prequantum line bundles from the Chern-Simons cocycle line bundles
using the method of  ~\cite{RSW}. 

We take up the case of an $SU(2)$ principal bundle $P$ and denote by $P^c$ when the group is complexified, i.e. $SL(2, \mathbb{C})$. Then  
 any  flat connection on $P^c$ is (upto gauge transformation) 
 is a $sl(2, \mathbb{C})$ connection of the form
\begin{eqnarray*}
B_{\lambda} &=& A + \lambda \Phi^{(1,0)} + \frac{1}{\lambda}\Phi^{(1,0)*} \\
&=& A + \lambda \Phi^{(1,0)} - \frac{1}{\lambda}\Phi^{(0,1)} 
\end{eqnarray*}
 where $|\lambda|^2 =1.$  This is because
$B_{\lambda}$ is of the form
$A + i \Psi$ where $A$ and $\Psi$ are unitary.  
This decomposition is always possible since ${\rm ad} P^c =  {\rm ad} P + i {\rm ad} P.$ Papers which use similar decomposition are ~\cite{Do}, ~\cite{C}. 
(Note $\Psi$ is unitary since $|\lambda|^2 =1$ and $\Phi = \Phi^{1,0} + \Phi^{0,1}$ is unitary). Flatness of $B_{\lambda}$ for all $\lambda \in S^1$ is equivalent to the fact that $(A, \Phi^{(1,0)})$ satisfy the self-duality equations (which is easy to check).
 Thus the moduli space of 
connections  $B_{\lambda}$ which are flat for all $\lambda$ is the moduli space of Hitchin systems, namely ${\mathcal M}$.

We consider now the Chern-Simons integral
$$Z = \int D {\bf A}  exp (i k {\rm CS} ({\bf A}))$$ where 
$${\rm CS} ({\bf A}) = 
\frac{1}{4 \pi} \int_{N} {\rm Tr} ({\bf A} \wedge d {\bf A} + \frac{2}{3} {\bf A} \wedge {\bf A}  \wedge {\bf A}).$$ Here $N$ is a $3$-manifold such that 
$\partial N = M,$ where $M$ is our original compact Riemann surface.

In what follows we will take ${\bf A} = {\bf B_{\lambda}}$ which is an 
extension of the $sl(2, \mathbb{C})$ connection $B_{\lambda}$ mentioned before  on $M$ to $N$. $\tilde{g}$ is an extenstion of the $SL(2, \mathbb{C})$ 
gauge transformation to $N$.
 As in ~\cite{RSW} we define the Chern-Simons cocycle to be
$$ \Theta( B_{\lambda}, g) = {\rm exp} i ({\rm CS} ({\bf B_{\lambda}}^{\tilde{g}}) - {\rm CS} ({\bf B_{\lambda}} ) ) $$ with which we define a line bundle on ${\mathcal M}$ (which is identified with flat $B_{\lambda}$ connections as mentioned before)  in what follows.

$${\mathcal L}_{\lambda} = {\mathcal M} \times_{\Theta} \mathbb{ C}$$
where there is a quotient by means of the equivalence relation:
$$(B_{\lambda}, z) \equiv ( B_{\lambda}^g, \Theta(B_{\lambda}, g) z)$$

As in ~\cite{RSW}, the curvature of this line bundle can be computed to be
$$F_{\lambda} (\tilde{\alpha}, \tilde{\beta}) = \frac{i}{2 \pi} {\rm Tr} \int_M \tilde{\alpha}\wedge \tilde{\beta} $$
where $$\tilde{\alpha} = \alpha^{1,0} + \alpha^{0,1} + \lambda \gamma^{1,0} - \frac{1}{\lambda} \gamma^{0,1},$$ $$\tilde{\beta} = \beta^{1,0} + \beta^{0,1} + \lambda \delta^{1,0} - \frac{1}{\lambda} \delta^{0,1}.$$ 

Thus, $F_{\lambda} = \frac{i}{2\pi} (\omega_1 + \lambda \omega_2 + \frac{1}{\lambda} \omega_3)$, where $\omega_1 (X, Y)  = \Omega (X, Y) =  \int_M {\rm Tr} (\alpha \wedge \beta - \gamma \wedge \delta), $ $\omega_2 (X, Y) =  \int_M {\rm Tr} (\alpha^{0,1} \wedge \delta^{1,0} + \gamma^{1,0} \wedge \beta^{0,1}),$ $\omega_3(X,Y) = - \int_M {\rm Tr} (\alpha^{1,0} \wedge \delta^{0,1} + \gamma^{0,1} \wedge \beta^{1,0})$. 
Thus  we obtain a whole $S^1$ worth of line bundles whose curvatures are 
parametrised by $\lambda.$

To construct  the prequantum line bundles ${\mathcal P}$, ${\mathcal E}$ and ${\mathcal N}$ from this family ${\mathcal L}_{\lambda}$ we 
note that 
if $ \lambda = \pm i$, $ F_{\lambda} = \frac{i}{2\pi} (\Omega \mp i Q_1) $ and 
if $\lambda = \pm 1$, $F_{\lambda} = \frac{i}{2\pi} (\Omega \mp i Q_2) $

Thus $\tau = {\mathcal L}_{i} \otimes {\mathcal L}_{-i} $ has curvature 
$\frac{i}{\pi} \Omega$. Thus $\tau $ and ${\mathcal P}$ have the same curvature. We say can that the Chern classes of these two line bundles are same as well when $H^{2} ({\mathcal M}, \mathbb{Z})$ has no torsion.  When the associated 
bundle $V$ to $P$ in ~\cite{H} is of 
rank two and degree odd , the Hitchin moduli 
space ${\mathcal M}$ is smooth and $H^2({\mathcal M}, \mathbb{Z})$ has no 
torsion \cite{H2}.  Since at least in this situation the Chern class cannot 
be torsion, curvature will 
 determine the Chern class, this line bundle $\tau$ is 
topologically equivalent to ${\mathcal P}$.      

Also, ${\mathcal L}_{i}^2 \otimes \tau^{-1}$ has curvature $\frac{1}{\pi}  {\mathcal Q}_1$ which is exact and thus since the moduli space has no torison the 
Chern class is zero and hence the bundle is trivial.
Since the curvature of ${\mathcal E} $ is $\frac{i}{\pi}  {\mathcal Q}_1$
it is also trivial. Since these are trivial bundles, they are isomorphic
and one can put $i$-times the connection on ${\mathcal L}_{i}^2 \otimes \tau^{-1}$ to get the connection on ${\mathcal E}$.

 Similarly, 
${\mathcal L}_1^2 \otimes \tau^{-1} $ has curvature $\frac{1}{\pi} Q_2$ which 
is exact and thus since the moduli space has no torsion  the Chern class is zero and hence the bundle is trivial. ${\mathcal N}$ has curvature  
$\frac{i}{\pi} Q_2$ and hence it is also trivial. Since these are trivial
 bundles they are isomorphic and one can put $i$ times the connection on  ${\mathcal L}_1^2 \otimes \tau^{-1} $ to get the connection on ${\mathcal N}$. 

Thus it is possible to get the prequantum line bundles from these cocycle
line bundles ${\mathcal L}_{\lambda}$ in the special case when the moduli space is smooth. In general,  the moduli  space is an orbifold, and perhaps there
 could be torsion in $H^2({\mathcal M}, \mathbb{Z})$. Then it is not clear that having the same curvature  would imply the lines bundles are topologically 
isomorphic.

{\bf Further work:} It would be interesting to see if these cocycle line bundles ${\mathcal L}_{\lambda} $ can be used to get some topological or geometrical invariants of $3$-manifolds, as in perhaps ~\cite{G}.

{\bf Acknowledgement:} I would like to thank Professor Nigel Hitchin,
Professor Jonathan Weitsman, Professor Ramadas and Professor Rajesh Gopakumar
  for the very useful discussions.

School of Mathematics, Harish Chandra Research Institute, Jhusi,
Allahabad, 211019, India. email: rkmn@mri.ernet.in

\end{document}